\documentclass[prd,aps,a4paper,twocolumn,showpacs,nofootinbib]{revtex4}

\usepackage{graphicx,psfrag}
\usepackage{mathrsfs}
\usepackage{amsmath,amsfonts,amssymb}
\usepackage{multirow}
\usepackage{comment}
\usepackage{bm}

\usepackage{graphicx}
\usepackage{psfrag}
\usepackage{epstopdf}
\usepackage{color}
\usepackage{mathrsfs}
\usepackage{comment}
\usepackage{multirow}
\usepackage{hyperref}
\usepackage{amsmath,amssymb,amsthm,wasysym}

\def\lm{{\ell m}}

\def\ha{{\hat{a}}}

\def\lm{{\ell m}}

\def\F{{\cal F}}

\def\ii{{\rm i}}

\def\ha{{\hat{a}}}

\newcommand{\be}{\begin{equation}}
\newcommand{\ee}{\end{equation}}

\definecolor{gray}{rgb}{0.5,0.5,0.5}
\definecolor{cyan}{rgb}{0,0.9,0.9}
\definecolor{orange}{rgb}{0.9,0.5,0}
\definecolor{magenta}{rgb}{1,0,1}
\definecolor{purple}{rgb}{0.8,0.4,0.8}
\definecolor{darkgreen}{rgb}{0,.6,0}
\definecolor{turquoise}{rgb}{0.25,0.88,0.82}

\begin{document}

\title{The antikick strikes back: recoil velocities for nearly-extremal\\ binary black hole mergers in the test-mass limit}

\author{Alessandro Nagar${}^1$, Enno Harms${}^2$, Sebastiano Bernuzzi${}^2$, An\i l Zengino\u{g}lu${}^{1}$}
\address{${}^1$Institut des Hautes Etudes Scientifiques, 91440 Bures-sur-Yvette, France}
\address{${}^2$Theoretical Physics Institute, University of Jena,
  07743 Jena, Germany}

\begin{abstract}
Gravitational waves emitted from a generic binary black-hole merger
carry away linear momentum anisotropically, resulting in a
gravitational recoil, or ``kick", of the center of mass. 
For certain merger configurations the time evolution of the magnitude of the 
kick velocity has a local maximum followed by a sudden drop. Perturbative 
studies of this ``antikick" in a limited range of black hole spins have found
that the antikick decreases for retrograde orbits as a function of
negative spin.  
We analyze this problem using a recently developed code to evolve
gravitational perturbations from a point-particle in Kerr spacetime
driven by an effective-one-body resummed radiation reaction force at
linear order in the mass ratio $\nu\ll 1$. 
Extending previous studies to nearly-extremal negative spins, we find that the
well-known decrease of the antikick is overturned and, instead of approaching
zero, the antikick increases again to reach $\Delta v/(c\nu^{2})=3.37\times10^{-3}$ 
for dimensionless spin $\ha=-0.9999$. The corresponding final kick velocity is $v_{end}/(c\nu^{2})=0.076$.
This result is connected to the nonadiabatic character of the emission of linear 
momentum during the plunge. We interpret it analytically by means of the 
\emph{quality factor}  of the flux to capture quantitatively the main properties 
of the kick velocity.
The use of such quality factor does not require trajectories nor horizon 
curvature distributions and should therefore be useful both in perturbation theory 
and numerical relativity.
\end{abstract}


\maketitle

\section{Introduction}
\label{sec:intro}

The anisotropic emission of gravitational radiation in coalescing black hole
binaries carries away linear momentum from the system, which results in a 
net recoil of the center of mass.
This gravitational recoil, or ``kick'', can be related to a delicate
and complicated interference between the gravitational wave (GW) multipoles. 
In the test-mass limit the recoil can be computed using
perturbative methods by modeling the small black hole as a
point-particle. Perturbative studies are crucial to study the
basic features of the interference pattern among different multipoles.
A detailed understanding of the recoil in the
perturbative regime is important not only for binaries with an extreme mass ratio, 
but also for comparable masses. As pointed out in Ref.~\cite{Nagar:2013sga} 
extrapolation from the test-mass result delivers quantitative agreement with numerical 
relativity for non-rotating black holes. Furthermore, for a rotating central 
black hole only the perturbative framework can systematically probe the extremal regime. 

Recoil computations in the test-mass limit were performed recently by two
groups using time domain calculations. 
The case with a non-rotating central black hole was studied in~\cite{Bernuzzi:2010ty} 
solving the Regge-Wheeler-Zerilli (RWZ) equations for gravitational metric perturbations. 
The case with a rotating central black hole was studied in~\cite{Sundararajan:2010sr} 
(SKH hereafter) solving the Teukolsky equation for gravitational curvature perturbations. 
The SKH analysis was limited to spin magnitudes $|\ha|\leq0.9$, 
where $\ha$ is the dimensionless angular momentum parameter. In
particular, SKH studied the drop in the time evolution of the recoil velocity, 
or ``antikick''~\cite{Baker:2006vn,Schnittman:2007ij}, as a function of
spin, and found that it is ``essentially non-existent" for large spin 
retrograde coalescences.

Building on recent progress in solving numerically the Teukolsky equation  
with a point-particle source in the time domain~\cite{Harms:2014dqa}, 
we revisit the SKH analysis and extend it to {\it nearly-extremal} spin values, particularly 
focusing on the retrograde case with spin parameters up to $\ha = - 0.9999$.
The extension of the parameter space reveals a new phenomenon: 
the antikick significantly reappears for $-1<\ha<-0.9$.
We explain this phenomenon by analyzing and relating the dynamics of the
plunge and the GW linear momentum flux. As noted long ago by Damour and 
Gopakumar~\cite{Damour:2006tr} (DG hereafter) the time-evolution of the 
recoil velocity (also for the comparable mass ratio case) and, in 
particular, the existence of an antikick can be directly connected to the 
nonadiabatic emission of linear momentum during the plunge.
Following DG, the behavior of the antikick as a function of $\ha$ 
is understood analytically and quantified in a ``quality factor'' $Q$
associated to the maximum of the GW linear momentum flux (Sec.~\ref{sec:kick}).
This understanding of the antikick relies on gauge-invariant notions and may
be a useful alternative to previous discussions that emphasize the 
trajectory~\cite{Price:2011fm, Price:2013paa} or curvature distributions 
on the horizon~\cite{Rezzolla:2010df}. 

To set the stage for our analysis, we discuss the dynamics of the system 
providing a quantitative measure for its nonadiabaticity (Sec.~\ref{sec:dyn}), 
and point out interesting properties of the GW linear
momentum flux (Sec.~\ref{sec:linflux}): as $\ha\to -1$, the linear momentum 
flux shows a characteristic, multi-peaked interference pattern that 
can be explained by the increased importance of the subdominant waveform
multipoles $0\leq m <\ell$ during the plunge~\cite{Harms:2014dqa}.
The behavior  of the maximal and final recoil velocity is discussed and analytically
explained in Sec.~\ref{sec:kick}. We examine the accuracy of our results in the
Appendix, including extremal positive spins, $+0.9\leq\ha\leq +0.9999$, 
that require special care.

We use geometric units $c=G=1$. The dynamics of the particle is 
obtained using a Hamiltonian formulation~\cite{Harms:2014dqa,Damour:2014sva} 
and expressed in dimensionless canonical variables.

\section{Dynamics: measuring nonadiabaticity}
\label{sec:dyn}

In the test-mass limit we model the black-hole binary system by a central
spinning black hole of mass $M$ and a nonspinning particle of mass
$\mu$, such that $\nu=\mu/M\ll 1$. Our test-mass calculations follow 
the method developed in~\cite{Nagar:2006xv,Bernuzzi:2010ty}, extended to the
Kerr background in~\cite{Harms:2014dqa}. The gravitational waveforms used to 
compute the flux of linear momentum are extracted at future null infinity with a
perturbative method based on the solution of the Teukolsky equation in
the time domain. The black hole spin is either aligned 
or anti-aligned with the orbital angular momentum. The relative dynamics is 
driven by an effective-one-body resummed analytic radiation 
reaction~\cite{Damour:2008gu,Pan:2010hz} at linear order in $\nu$. 
For simplicity, we do not include horizon absorption~\cite{Nagar:2011aa,Taracchini:2013wfa},
so that the radiation reaction only incorporates the angular momentum
flux emitted to infinity, following~\cite{Harms:2014dqa}.
Since our radiation reaction is certainly inaccurate as $\ha\to 1$
(because of both the absence of horizon absorption and  
the lack of higher-order spin-dependent terms in the resummed flux
at infinity~\cite{Harms:2014dqa,Taracchini:2014zpa}) our results for large, 
positive spin may be partly affected by systematic uncertainties. For this reason, 
we discuss in the main text only the spin range $-0.9999\leq \ha \leq +0.9$, while 
the more challenging\footnote{Note that by ``challenging'' we refer here to the limits of 
the radiation reaction model and {\it not} to the solution of the Teukolsky equation using
the methods of Ref.~\cite{Harms:2013ib,Harms:2014dqa}. The inclusion of the higher-order 
post-Newtonian information of Ref.~\cite{Shah:2014tka} in resummed form (not available at the moment) 
in the radiation reaction would certainly allow us to improve our approach.} 
regime $+0.9<\ha \leq +0.9999$ is discussed separately in Appendix~\ref{sec:acc_fit}.
Our main new findings are in the regime $\ha\to-1$, where the analytic 
radiation reaction is robust. We work with mass ratio $\nu=10^{-3}$; the spin 
configurations we consider are listed in Table~4 of~\cite{Harms:2014dqa}.

\begin{figure}[t]
\centering
 \includegraphics[width=0.5\textwidth]{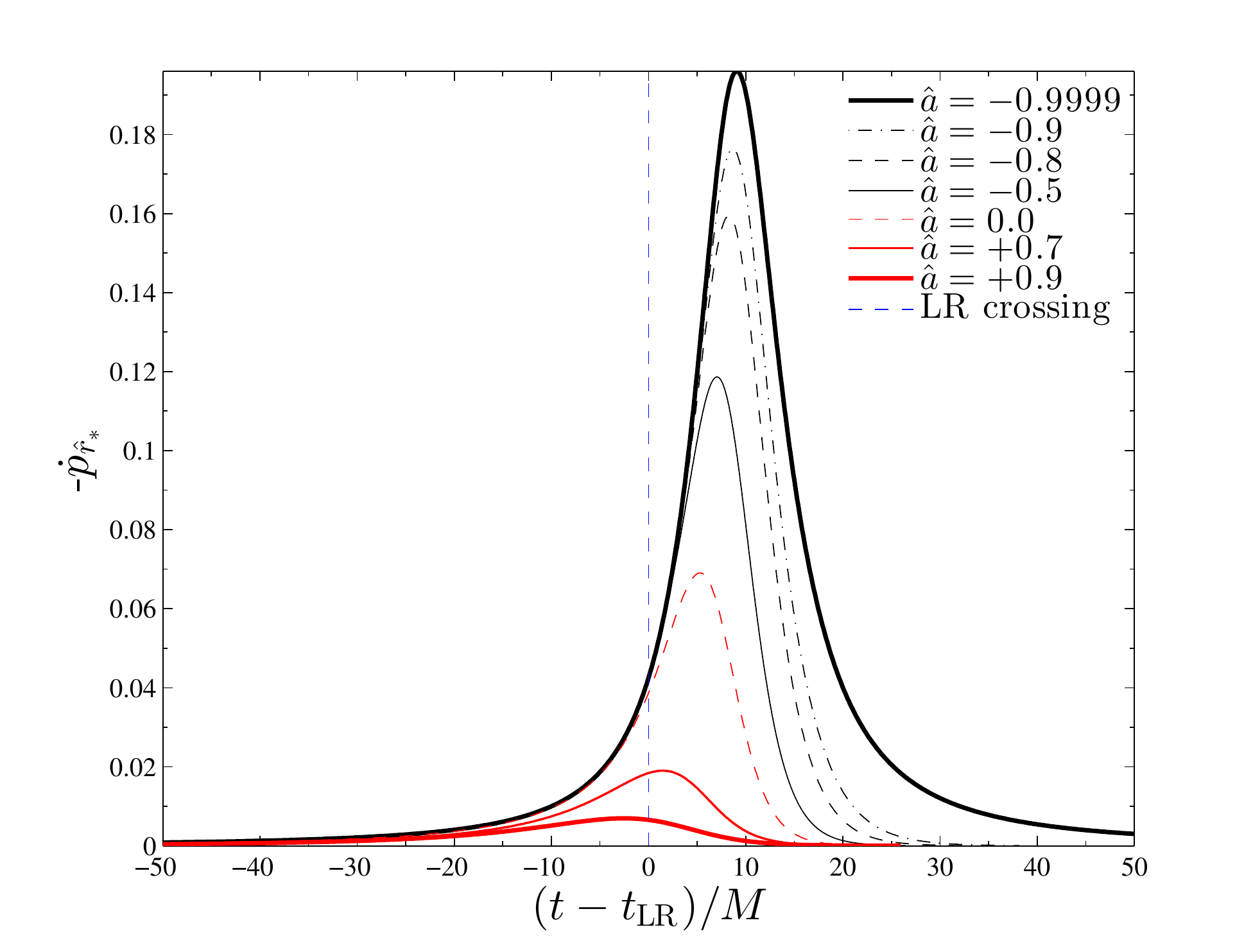}
    \caption{
    Time evolution of $-\dot{p}_{r_{*}}$: the characteristic time-scale of the 
    curve $\tau_{\dot{p}_{r_{*}}}^{max}$ (Eq.~\eqref{eq:tau_max}) is a measure of the adiabaticity 
    of the plunge. One finds that $\tau_{\dot{p}_{r_{*}}}^{max}$ decreases from $\ha=0.9$ to 
    $\ha=-0.57$ (see also Table \ref{tab:finalV}), but then increases again. This is consistent 
    with the increase of the quality factor $Q$ (indicating adiabaticity in the linear momentum flux)
    and the related peculiar time evolution of the recoil velocity as $\ha\to -1$ (see Fig.~\ref{fig:vkick}).
    Note that $t_{\rm LR}$ indicates the light-ring crossing time.}
 \label{fig:dotprstar}
\end{figure}

The relative dynamics is started from post-circular initial data~\cite{Buonanno:2000ef,Nagar:2006xv}
and driven from inspiral to plunge by the radiation reaction. 
The transition from quasi-circular inspiral to plunge
depends on the spin-orbit coupling 
between the particle's angular momentum and the black-hole's spin through
the Hamiltonian. It can be 
slowly-varying and adiabatic   
(spin aligned with particle's angular momentum, the last-stable-orbit (LSO) moves
towards the horizon)  
or quickly-varying and nonadiabatic (spin anti-aligned with particle's angular
momentum, the LSO moves away from 
the horizon). The net GW emission of linear momentum and the final value
of the recoil velocity 
can be connected to the nonadiabatic part of the
dynamics~\cite{Damour:2006tr}.
(A similar argument has also 
been discussed recently in Refs.~\cite{Price:2011fm,Price:2013paa}).
In the following, we introduce a quantitative measure of this nonadiabaticity 
in the plunge phase.

Consider the time derivative of the radial momentum 
in a tortoise coordinate  $-\dot{p}_{r_*}$ (changed sign for clarity; see
Ref.~\cite{Harms:2014dqa} for the precise definition). As shown in
Ref.~\cite{Harms:2014dqa} (see Fig.~15 there), $-p_{r_*}$ is a monotonic
function of time: it grows during the plunge 
attaining a finite maximum at the horizon. Its time derivative has a bell shape 
as displayed in Fig.~\ref{fig:dotprstar} for a few representative values of
$\ha$. For convenience of comparison, the plot is done versus 
$t-t_{\rm LR}$, where $t_{\rm LR}$ is the light-ring crossing time 
defined by $r_{\rm LR}\equiv  r(t_{\rm LR})$ and 
$r_{\rm LR}=2\left[1+\cos\left(\frac{2}{3}\arccos(-\ha)\right)\right]$.

The spin-orbit interaction is repulsive for prograde orbits and attractive
for retrograde orbits.
Consistently, the distribution of $-\dot{p}_{r_{*}}$ is wider as $\ha\to +1$
(slowly-varying, adiabatic plunge dynamics) 
and narrower as $\ha\to -1$ (quickly-varying, nonadiabatic plunge dynamics). 
To quantify the spin-dependence of the width of the curve, 
we define its characteristic variation time
\be
\label{eq:tau_max}
\tau_{\dot{p}_{r_{*}}}^{max} = -\dfrac{\dot{p}_{r_{*}}}{\dddot{p}_{r_{*}}}\vert_{t=t_{max}^{\dot{p}_{r_{*}}}} \ ,
\ee 
where $t_{max}^{\dot{p}_{r_{*}}}$ corresponds to the peak of $-\dot{p}_{r_{*}}$. The values of $\tau_{\dot{p}_{r_{*}}}^{max}$ are
listed in Table~\ref{tab:finalV}. 
Note that $\tau_{\dot{p}_{r_{*}}}^{max}$ {\it is not}  
monotonically decreasing when the spin decreases from positive
to nearly-extremal negative values (it is not possible to
deduce this from the plot). 
On the contrary, $\tau_{\dot{p}_{r_{*}}}^{max}$ 
attains a minimum for $\ha \sim -0.57$ 
and grows again as $\ha\to -1$ (though to smaller values), 
indicating that the dynamics becomes slightly more adiabatic 
again\footnote{Since $p_{r*}$  attains values larger 
than 1 around the light-ring crossing, as seen in Fig.~15 of 
Ref.~\cite{Harms:2014dqa}, one may have some nonnegligible contribution of
the radial part of the radiation reaction $\F_{r_{*}}$ as $\ha\to -1$. 
This term is not included  in the dynamics because of the current lack of a robust 
resummation strategy for the post-Newtonian expanded results of Ref.~\cite{Bini:2012ji}. 
Still, we have verified that the inclusion of the leading order term 
$\F_{r_{*}}=-\frac{5}{3}\frac{p_{r_{*}}}{p_{\phi}}\F_{\phi}$ (here $p_{\phi}$ is the 
mechanical angular momentum and $\F_{\phi}$ its resummed loss~\cite{Harms:2014dqa}) 
does not have any visible effect on the plunge dynamics. This makes us confident
that indirect plunges are essentially geodetic.}.
Such a simple quantitative characterization of the plunge is helpful in interpreting 
the following analysis of the linear momentum flux and the recoil velocity.

\section{The GW linear momentum flux}
\label{sec:linflux}

\begin{figure}[t]
\centering
 \includegraphics[width=0.4\textwidth]{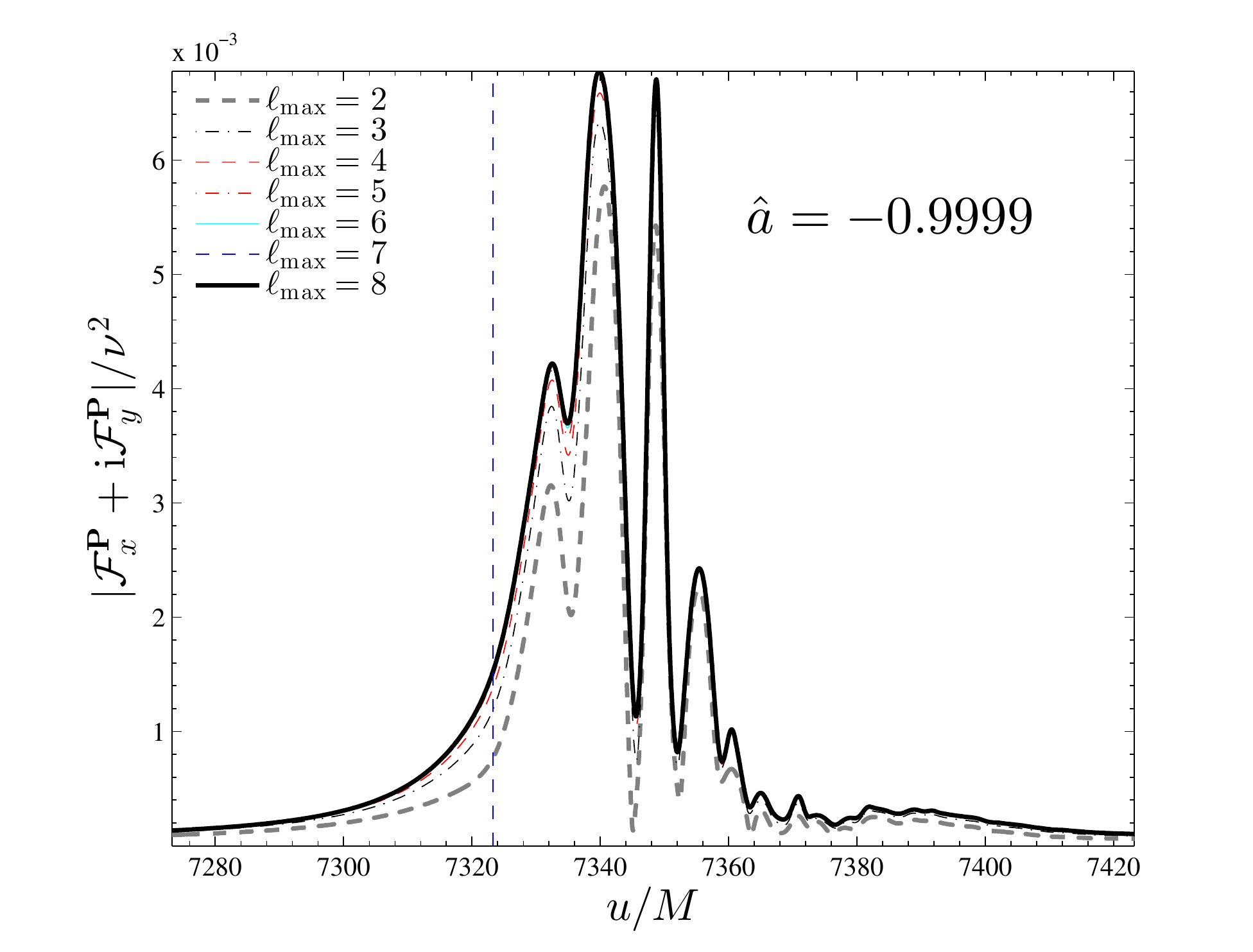}\\
 \includegraphics[width=0.4\textwidth]{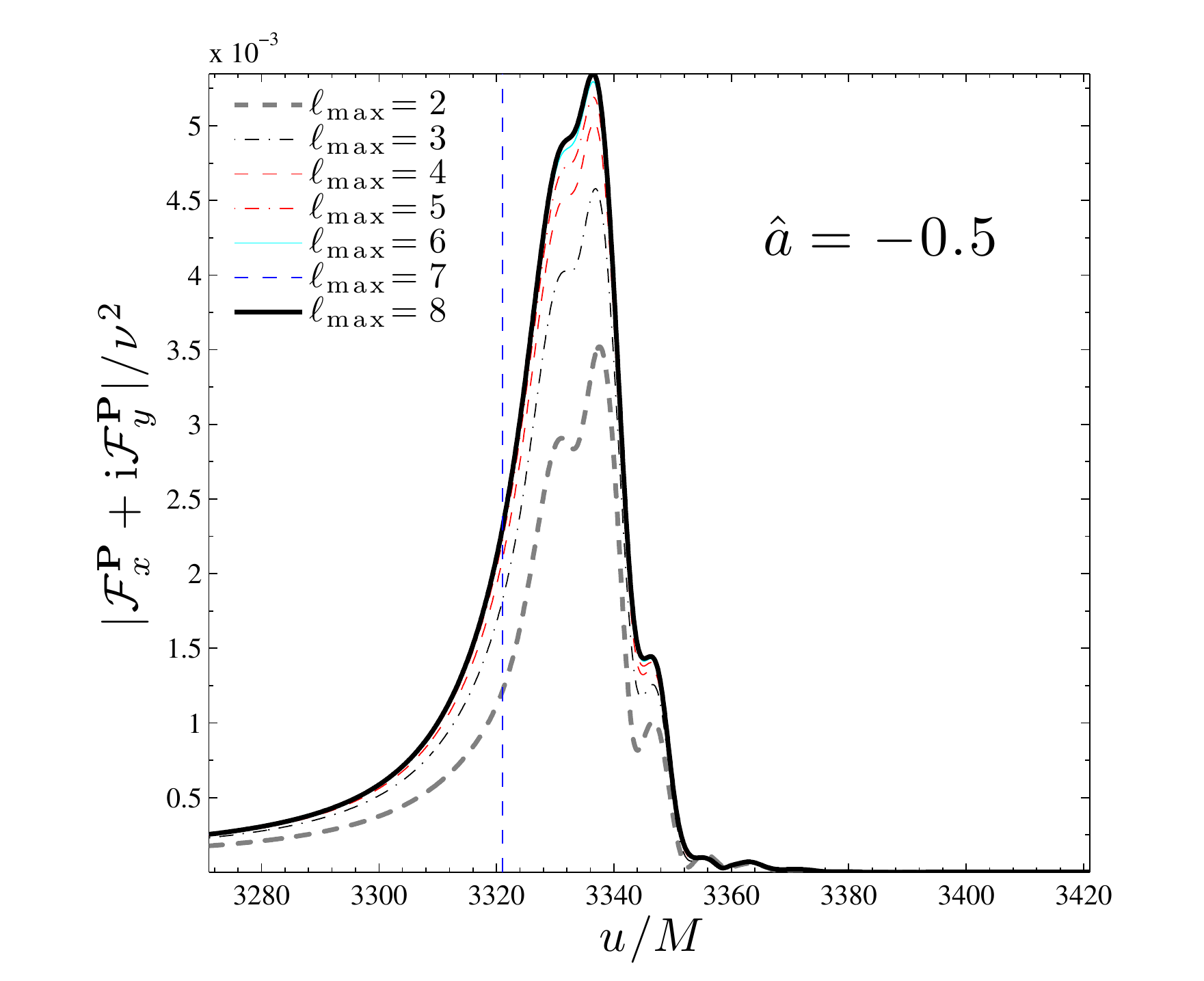}\\
 \includegraphics[width=0.4\textwidth]{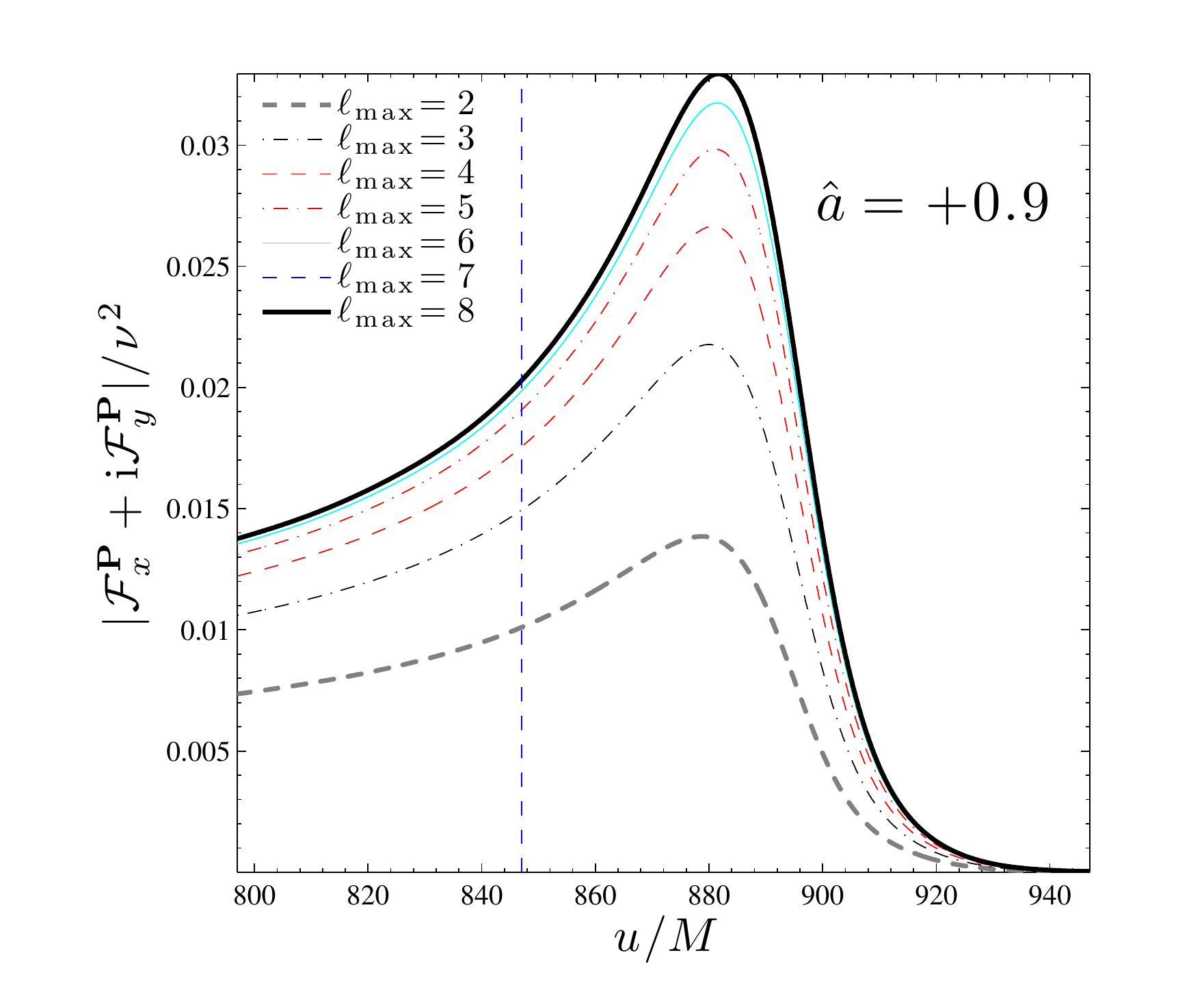}
    \caption{ \label{fig:Pflux} Modulus of the linear momentum flux
     for three representative values of $\ha$. As $\ha\to -1$, the emission
     of linear momentum occurs in a shorter time. The interference pattern 
     seen for  $\ha=-0.9999$ is determined by the increased importance of 
     the subdominant waveform multipoles with $0\leq m < \ell$ when $\ha\to -1$ 
     (as noted in Ref.~\cite{Harms:2014dqa}) around merger (defined as the peak 
     of $|\Psi_{22}|$, dashed vertical lines).}  
\end{figure}

\begin{figure}[t]
\centering
 \includegraphics[width=0.4\textwidth]{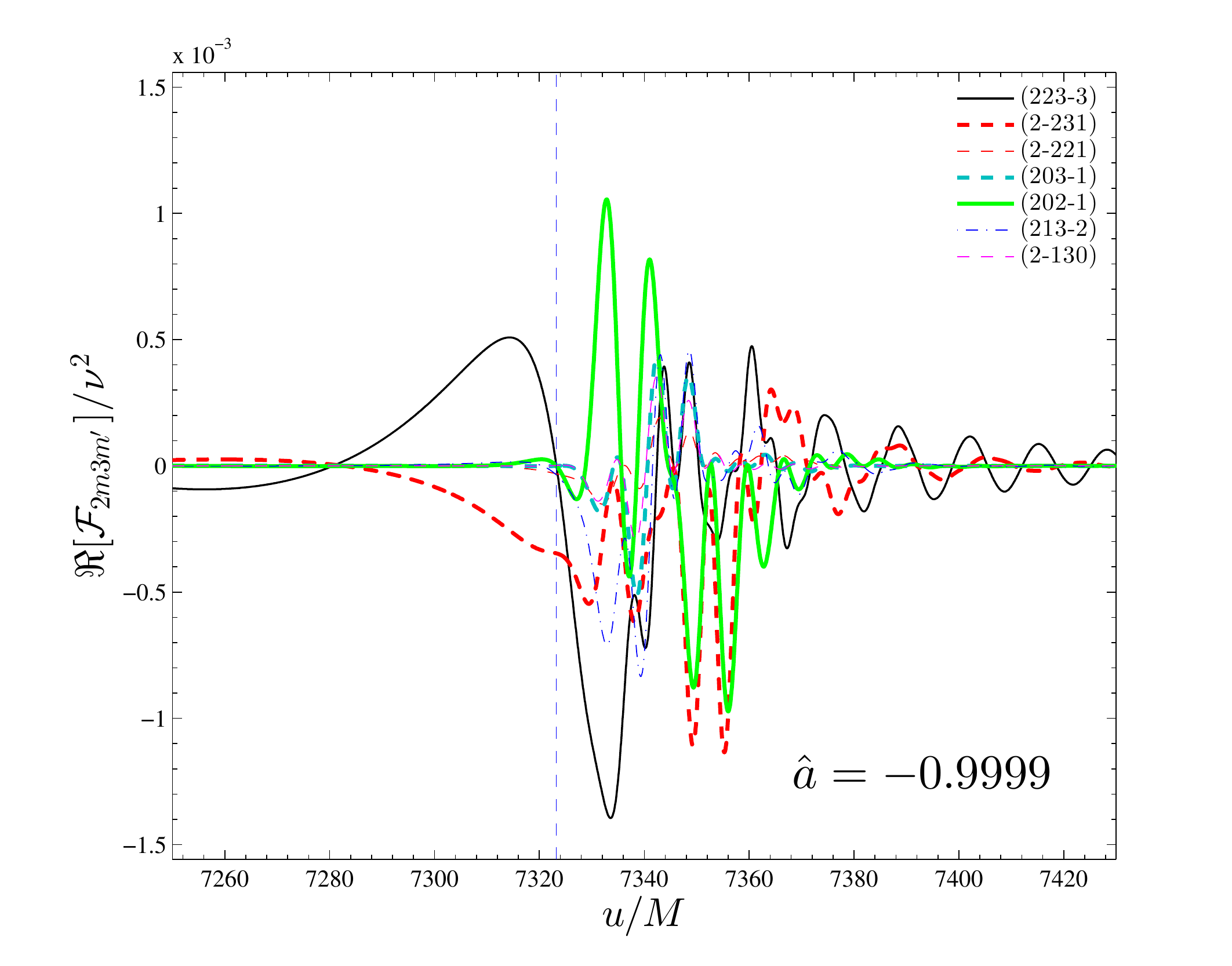}\\
 \includegraphics[width=0.4\textwidth]{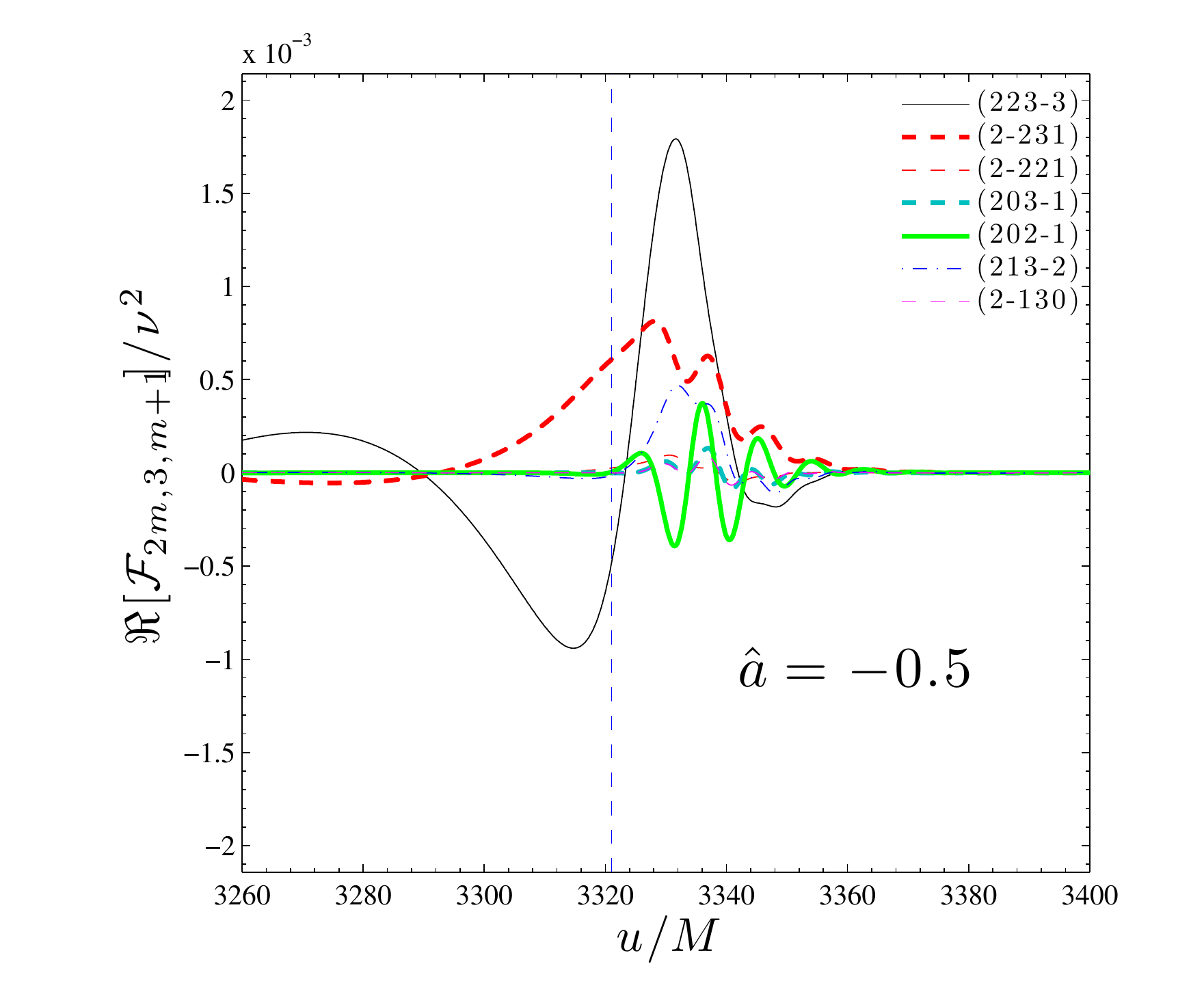}
 \includegraphics[width=0.4\textwidth]{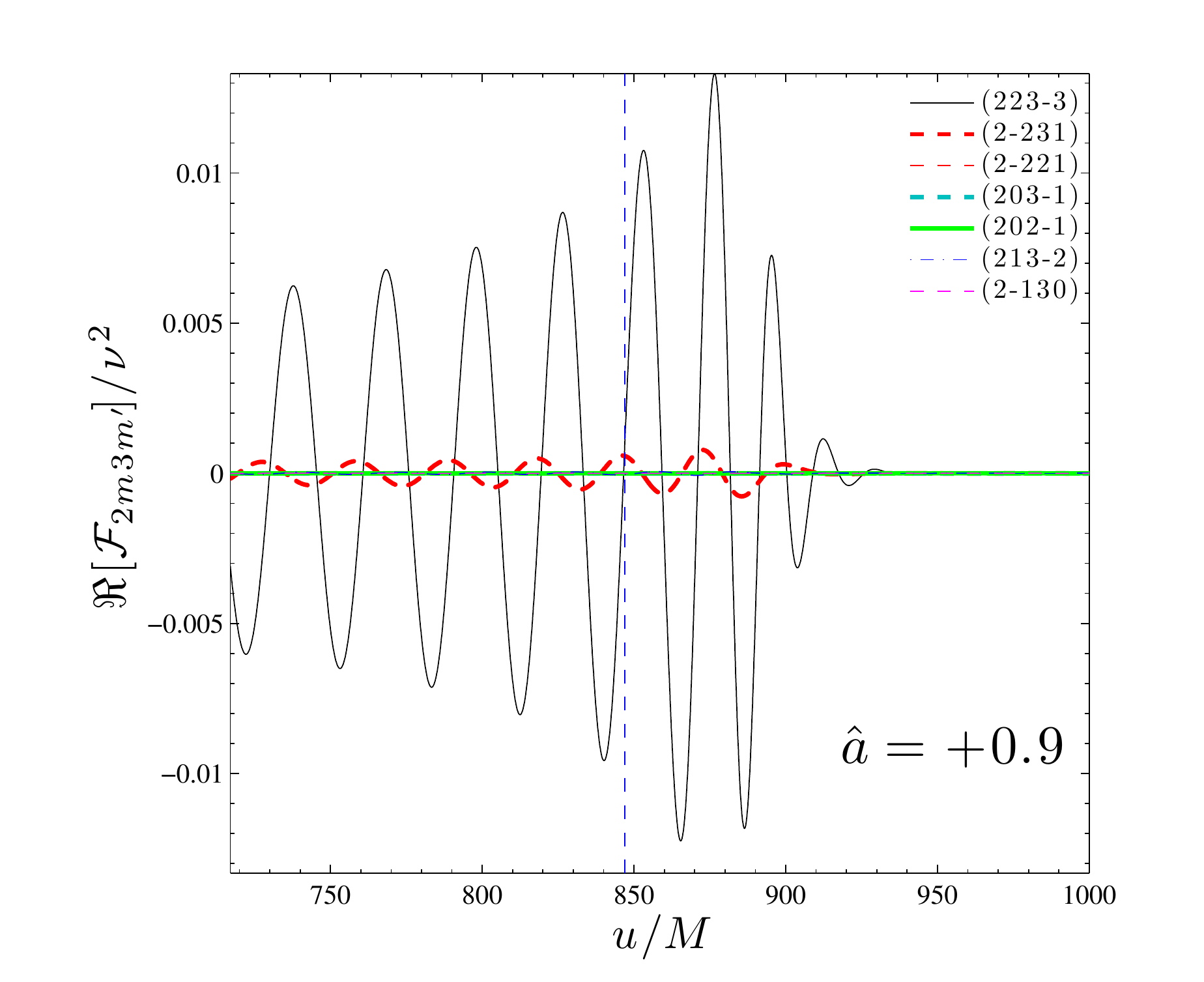}
    \caption{Comparing the real part of the various terms entering the leading contribution
     ${\cal F}_{2}^{\bf P}$ to the linear momentum flux, Eq.~\eqref{eq:F2}. One sees that for $\ha=-0.9999$ 
     all terms in Eqs.~\eqref{f1}-\eqref{f7} have comparable magnitudes around merger (marked by the vertical line). 
     This prompts the interference pattern seen in the corresponding
     modulus in Fig.~\ref{fig:Pflux}.} 
    \label{fig:ReF}
\end{figure}

Let us now analyze the GW linear momentum flux. We will see how the emission 
of linear momentum closely mirrors the plunge dynamics. Notably, the analysis 
of the flux (a gauge invariant quantity) is independent of having at hand a 
description of the dynamics and therefore can be directly applied 
to investigate also numerical relativity data. 

In our simulations the GW linear momentum is emitted in the
equatorial $xy$-plane because we consider equatorial orbits (the
black hole spin is either aligned or antialigned with the orbital
angular momentum). Working with RWZ-normalized variables
$\Psi_{\lm}^{(\epsilon)}$ the GW linear momentum flux reads
\begin{align}
\label{eq:Pflux}
\F_x^{\bf P} + \ii \F_y^{\bf P}= \sum_{\ell=2}^{\ell_{max}}\F^{\bf
  P}_\ell=\dfrac{1}{8\pi}\sum_{\ell=2}^{\ell_{max}}\sum_{m=-\ell}^\ell\ii
\bigg[a_{\lm}\dot{\Psi}_\lm^{(0)}\dot{\Psi}^{(1)*}_{\ell,m+1}\nonumber\\ 
+b_\lm\sum_{\epsilon=0,1} \dot{\Psi}^{(\epsilon)}_\lm\dot{\Psi}^{(\epsilon)*}_{\ell+1,m+1}\bigg],
\end{align}
where the numerical coefficients $(a_\lm,b_\lm)>0$ are given in
Eqs.~(16)-(17) of~\cite{Bernuzzi:2010ty}, $\epsilon$ is the parity of
$(\ell+m)$, and $\Psi^*_{\lm} = (-1)^m\Psi_{\ell,-m}$. Note that for each value of $\ell$
the contribution $\F^{\bf P}_\ell$ involves all $\ell$ and $\ell+1$ waveform multipoles 
(e.g., for $\ell=2$ one deals with 7 waveform multipoles). Since we extracted gravitational 
wave multipoles up to $\ell_{max}=8$, we do not include $\ell=9$ modes in  $\F^{\bf P}_8$.

Figure~\ref{fig:Pflux} shows the flux of linear momentum as a function 
of the retarded time $u$ (cf.~\cite{Harms:2014dqa}) 
for $\ha=-0.9999$ (top), $\ha=-0.5$ (middle) and $\ha=+0.9$ (bottom). 
Each labeled line on the plot corresponds to the sum $\sum_{\ell=2}^{\ell_{max}}\F^{\bf P}_\ell$ 
in Eq.~\eqref{eq:Pflux} up to the indicated $\ell_{max}$. The vertical dashed line indicates
the ``merger time'' $u_{\rm mrg}$, defined as the time of the peak of $|\Psi_{22}|$.
To relate these figures with Fig.~\ref{fig:dotprstar}, as $\ha\to -1$ one has 
$u_{\rm mrg}\approx t_{\rm LR}$, while as $\ha\to 1$ one progressively gets $u_{\rm mrg}< t_{\rm LR}$.
The precise quantitative information is collected in Table~4 of~\cite{Harms:2014dqa}: one has 
$t_{\rm LR}=7321.7$ for $\ha=-0.9999$, $t_{\rm LR}=3321.3$ for $\ha=-0.5$ and $t_{\rm LR}=883.6$
for $\ha=+0.9$ [the corresponding last-stable-orbit (LSO) crossing times are 6858.3, 2980.4 and 820.7].

Comparing the three plots in Fig.~\ref{fig:Pflux} 
one can directly extract that as $\ha\to -1$: (i) the emission of linear momentum appears 
more localized in time
(the three time-axes show an equally-sized range of $\sim140M$)
, i.e.~it becomes an impulsive phenomenon; 
(ii) the simple single-peak structure is replaced by a complicated interference
pattern with several peaks of different amplitude and width.

This phenomenon mirrors strong destructive interference~\footnote{
  Although mode mixing is expected in the rotating Kerr background,
  here the interference phenomenon is of different physical
  origin. First, such interference is present already in the nonrotating
  background, e.g.~\cite{Nagar:2013sga}. Second, our discussion on a
  rotating background could be formulated only in terms of azimuthal
  $m$-modes which are an appropriate basis. Note, however, that 
  we stick to the full spin-weighted spherical harmonics decomposition 
  since in our setup the flux calculation in terms of $m$-modes only is
  technically more involved
  due to the coupling between $m$ and $m+1$ in
  Eq.~\eqref{eq:Pflux} (two different simulations). } 
effects between the various terms entering
Eq.~\eqref{eq:Pflux}. Such effect is maximal as $\ha\to -1$ and progressively
less apparent as $\ha$ increases. It can be explained (see below) by the magnification of 
the subdominant $0\leq m < \ell$ modes during the late plunge and merger as $\ha\to -1$.
Since it is present already in the leading-order $\F_{2}^{\bf P}$ term 
(dashed line in the bottom panel of Fig.~\ref{fig:Pflux}) it can be quantitatively understood 
by analyzing the behavior of only this contribution as a function of the black hole spin. 

Setting $\ell_{max}=2$ the corresponding GW linear momentum flux $\F_{2}^{\bf P}$ is built from 
the interference of the following seven terms, involving all $\ell=2$ and $\ell=3$ multipoles:
\begin{align}
\label{eq:F2}
{\F}_{2}^{\bf P}&=\F_{223-3}+\F_{2-231}+\F_{2-221}\nonumber\\
                                             &+\F_{202-1}+\F_{203-1}+\F_{213-2}+\F_{2-130} \ .
\end{align}
The $\F_{l m \ell' m'}$ are obtained  
from Eq.~\eqref{eq:Pflux} and read explicitly
\begin{subequations}
\label{f}
\begin{align}
\label{f1}
\F_{223-3}   & =\dfrac{5}{\pi}\sqrt{\dfrac{6}{7}}\dot{\Psi}_{22}\dot{\Psi}_{3-3},\\
\label{f2}
\F_{2-221} & = \dfrac{2\ii}{\pi}\dot{\Psi}_{2-2}\dot{\Psi}_{21},\\
\label{f3}
\F_{2-231} & = \dfrac{1}{\pi}\sqrt{\dfrac{10}{7}}\dot{\Psi}_{2-2}\dot{\Psi}_{31},\\
\label{f4}
\F_{202-1}&=\dfrac{\ii}{\pi}\sqrt{6}\,\dot{\Psi}_{20}\dot{\Psi}_{2-1},\\
\label{f5}
\F_{203-1} &=\dfrac{2}{\pi}\sqrt{\dfrac{15}{7}}\dot{\Psi}_{20}\dot{\Psi}_{3-1},\\
\label{f6}
\F_{213-2}&=-\dfrac{1}{\pi}\dfrac{10}{\sqrt{7}}\,\dot{\Psi}_{21}\dot{\Psi}_{3-2},\\
\label{f7}
\F_{2-130}&= -\dfrac{1}{\pi}\sqrt{\dfrac{30}{7}}\dot{\Psi}_{2-1}\dot{\Psi}_{30} \ ,
\end{align}
\end{subequations}
when using $\Psi^*_{\lm} = (-1)^m\Psi_{\ell,-m}$.

References~\cite{Harms:2014dqa,Taracchini:2014zpa} pointed out that the breakdown of the circularity
during the plunge as $\ha\to -1$ (see Fig.~15 in Ref.~\cite{Harms:2014dqa}) makes 
each multipolar waveform amplitude higher and sharper around their peak 
(which occurs near merger).
In particular for $0\leq m < \ell$ the peaks get amplified to values comparable to that 
of the leading  $\ell=m=2$ mode (the effect is particularly striking for the $m=0$ modes). 
This phenomenon occurs on the short time scale of the plunge and thus also
yields a magnification of the $\dot{\Psi}_\lm$'s. One can then understand how the 
spin-dependence of the  various $\F_{l m \ell' m'}$ terms in Eqs.~\eqref{f} can prompt 
complicated interference patterns via Eq.~\eqref{eq:F2}.
To illustrate how this works in practice,  Fig.~\ref{fig:ReF} compares the real part of the seven 
partial contributions given by Eq.~\eqref{f} for spins $\ha\in\{-0.9999,-0.5,+0.9\}$. 
For $\ha=-0.9999$ all terms in Eq.~\eqref{f} are comparable. One sees that
$\F_{202-1}$ and $\F_{2-231}$ are approximately in phase among themselves and 
in phase opposition to $\F_{213-2}$ and $\F_{223-3}$. When taking the modulus
of the sum of all these contributions one understands the origin of the minima
in the modulus of Fig.~\ref{fig:Pflux}. Notably, the times of the minima in Fig.~\ref{fig:Pflux} 
correspond to the minima of $\F_{202-1}$ and $\F_{2-231}$, indicating that the interference
pattern of the linear momentum flux reflects the enhancement of $(\Psi_{20},\Psi_{21},\Psi_{31})$. 
This is driven by the next-to-quasi-circular corrections to the waveform, which are enhanced
for the mainly radial indirect plunges.

By contrast, when $\ha=+0.9$,  $\F_{223-3}$ is much larger than the other terms, e.g.
$\F_{2-231}$ and $\F_{202-1}$ do not contribute significantly. 
The negligible value of $\dot{\Psi}_{31}$ with respect to $\dot{\Psi}_{2-2}$ essentially 
removes the complicated behavior that one finds in $\F_{2-231}$ as $\ha\to -1$, 
and this contribution is just dominated by the $\dot{\Psi}_{2-2}$ mode. 
Note in the bottom panel of Fig.~\ref{fig:ReF} how the red and black lines
are dephased by $\pi/2$, consistent with the dephasing 
due to complex conjugation.

Finally, focusing on the case $\ha=-0.9999$ for definitess, we note that 
the emission of linear momentum predominantly occurs on the time interval $(7320,7360)$
around the largest peak of $|\F^{\bf P}|$; the interval is approximately the same where
$-\dot{p}_{r_{*}}$ is significantly different from zero ($-\dot{p}_{r_{*}}$
peaks at $t^{p_{r_{*}}}_{max}\approx 7331$). This supports the understanding 
that it is the time variation of $p_{r_{*}}$ that is pumping up (the time-derivatives of) 
the gravitational waveform around the light-ring crossing to generate the narrow 
burst of linear momentum. For this value of the spin, we also note that
the rather shallow  peak of the flux around $u/M\sim7390$ is essentially driven by the 
quasi-normal-mode excitation. For $\ha=-0.9999$ the modes are long-lasting,
which explains why this peak is so shallow (see also the top panel of Fig.~\ref{fig:ReF}).
The same feature, with the same explanation, is seen also for $\ha=-0.5$, though
it is absent for $\ha=+0.9$. We postpone to future work a detailed analysis of
the QNMs-driven features of the linear momentum flux.

\section{Kick and antikick}
\label{sec:kick}

\begin{figure}[t]
\centering
 \includegraphics[width=0.5\textwidth]{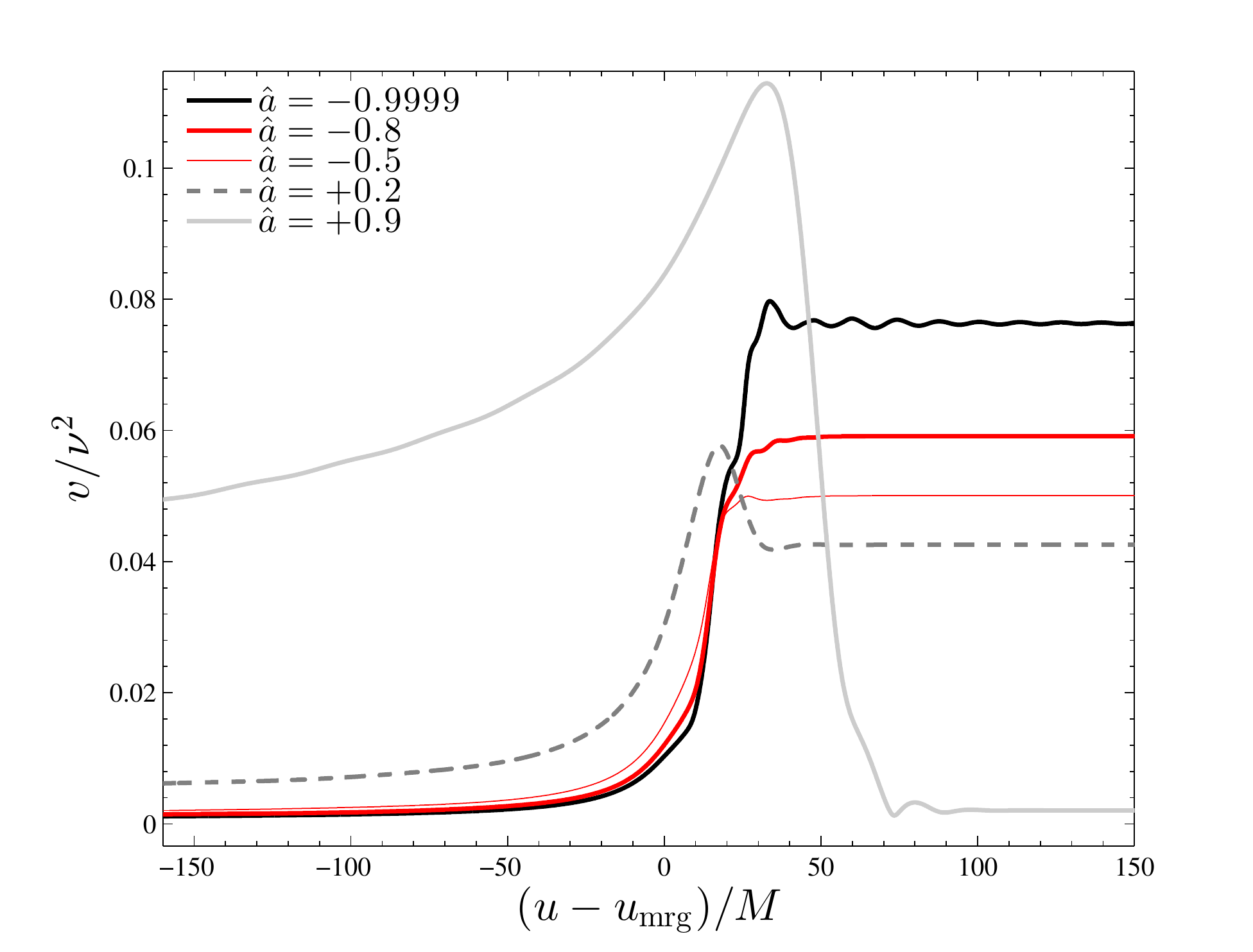}
    \caption{Time evolution of the recoil velocity for various black
      hole spin parameters $\ha$.
    The large antikick present for positive values of $\ha$ is progressively absorbed until it disappears
    when $-0.9\lesssim\ha\lesssim -0.5$. Suprisingly, for nearly-extremal negative spins it progressively 
    reappears due to a slight increase of the adiabatic character of the plunge dynamics. We use 
    the peak of $|\Psi_{22}|$ as the merger time $u_{\rm mrg}$.}
 \label{fig:vkick}
\end{figure}

\begin{table}[t]
 \caption{\label{tab:finalV} From left to right the columns report:
 the magnitude of the final and the maximal recoil velocities, 
 $v_{end}/\nu^{2}$ and $v_{max}/\nu^{2}$; the magnitude of the antikick $\Delta v/\nu^{2}$: 
  for  $-0.9\leq \ha \leq -0.5$ no significant antikick is observed; 
  the quality factor $Q$ associated with the maximum of the amplitude 
  of the linear momentum flux, as an indicator of the adiabaticity of the 
  emission of linear momentum. The larger is $Q$ the more adiabatic
   is the emission process, the larger is the antikick; the characteristic time
   scale $\tau_{\dot{p}_{r_{*}}}^{max}$ of $-\dot{p}_{r_{*}}$ (see Eq.~\eqref{eq:tau_max}),
   as a complementary indicator of the adiabaticity of the dynamics;
   the approximate analytic calculation of the kick velocity, $v_{end}^{anal}/\nu^{2}$ of
   Eq.~\eqref{eq:vend}. Minima of $\Delta v/\nu^{2},Q,\tau_{\dot{p}_{r_{*}}}^{max}$ are 
   printed in boldface.} 
     \centering
  \begin{ruledtabular}
  \begin{tabular}{l||ccc|ccc}
    $\ha$ & $v_{max}/\nu^{2}$ & $v_{end}/\nu^{2}$  &
    $\Delta v /\nu^{2}$ &  $Q$ & $\tau^{max}_{\dot{p}_{r_{*}}}$ & $v_{end}^{anal}/\nu^{2}$\\
    \hline 
    \hline   
    -0.9999 & 0.07972 & 0.07634 & 3.377e-03 & 1.0060 & 3.8436 & 0.04060 \\
    -0.9990 & 0.07967 & 0.07637 & 3.303e-03 & 1.0065 & 3.8411 & 0.04091 \\
    -0.9950 & 0.07884 & 0.07587 & 2.972e-03 & 0.9942 & 3.8302 & 0.04052 \\
    -0.9900 & 0.07798 & 0.07539 & 2.589e-03 & 0.9639 & 3.8171 & 0.04050 \\
    -0.9800 & 0.07571 & 0.07383 & 1.883e-03 & 0.9518 & 3.7924 & 0.04017 \\
    -0.9700 & 0.07452 & 0.07320 & 1.326e-03 & 0.9356 & 3.7696 & 0.03996 \\
    -0.9500 & 0.07093 & 0.07040 & 5.264e-04 & 0.9015 & 3.7292 & 0.03942 \\
    -0.9000 & 0.06545 & 0.06539 & 5.589e-05 & 0.8663 & 3.6508 & 0.03855 \\
    -0.8000 & 0.05910 & 0.05909 & 9.332e-06 & {\bf0.8378} & 3.5570 & 0.03807 \\
    -0.7000 & 0.05501 & 0.05501 & 8.223e-07 & 0.8402 & 3.5123 & 0.03910 \\
    -0.6000 & 0.05183 & 0.05183 & 1.915e-08 & 0.8650 & {\bf3.4977} & 0.04189 \\
    -0.5000 & 0.05003 & 0.05003 & {\bf2.289e-09} & 0.9024 & 3.5044 & 0.04765 \\
    -0.4400 & 0.04914 & 0.04879 & 3.485e-04 & 0.9491 & 3.5167 & 0.05318 \\
    -0.4000 & 0.04948 & 0.04882 & 6.618e-04 & 1.0038 & 3.5280 & 0.05801 \\
    -0.3000 & 0.04913 & 0.04766 & 1.479e-03 & 1.9191 & 3.5667 & 0.09562 \\
    -0.2000 & 0.04981 & 0.04658 & 3.224e-03 & 1.4625 & 3.6198 & 0.09148 \\
    -0.1000 & 0.05060 & 0.04534 & 5.266e-03 & 1.4011 & 3.6878 & 0.07821 \\
    0.0000  & 0.05319 & 0.04530 & 7.892e-03 & 1.4364 & 3.7722 & 0.07029 \\
    0.1000  & 0.05471 & 0.04377 & 1.094e-02 & 1.5086 & 3.8755 & 0.06279 \\
    0.2000  & 0.05771 & 0.04252 & 1.519e-02 & 1.6045 & 4.0019 & 0.05655 \\
    0.3000  & 0.06105 & 0.04053 & 2.052e-02 & 1.7207 & 4.1580 & 0.05116 \\
    0.4000  & 0.06606 & 0.03822 & 2.785e-02 & 1.8678 & 4.3534 & 0.04578 \\
    0.5000  & 0.07131 & 0.03398 & 3.733e-02 & 2.0643 & 4.6049 & 0.03887 \\
    0.6000  & 0.07796 & 0.02831 & 4.965e-02 & 2.3413 & 4.9426 & 0.02766 \\
    0.7000  & 0.08719 & 0.02056 & 6.663e-02 & 2.7528 & 5.4289 & 0.01406 \\
    0.8000  & 0.09919 & 0.01085 & 8.835e-02 & 3.5249 & 6.2242 & 0.00431 \\
    0.9000  & 0.11293 & 0.00206 & 1.109e-01 & 5.3834 & 7.8682 & 0.00031 \\
    \hline
  \end{tabular} 
  \end{ruledtabular}
\end{table}

\begin{figure}[t]
  \centering
   \includegraphics[width=0.45\textwidth]{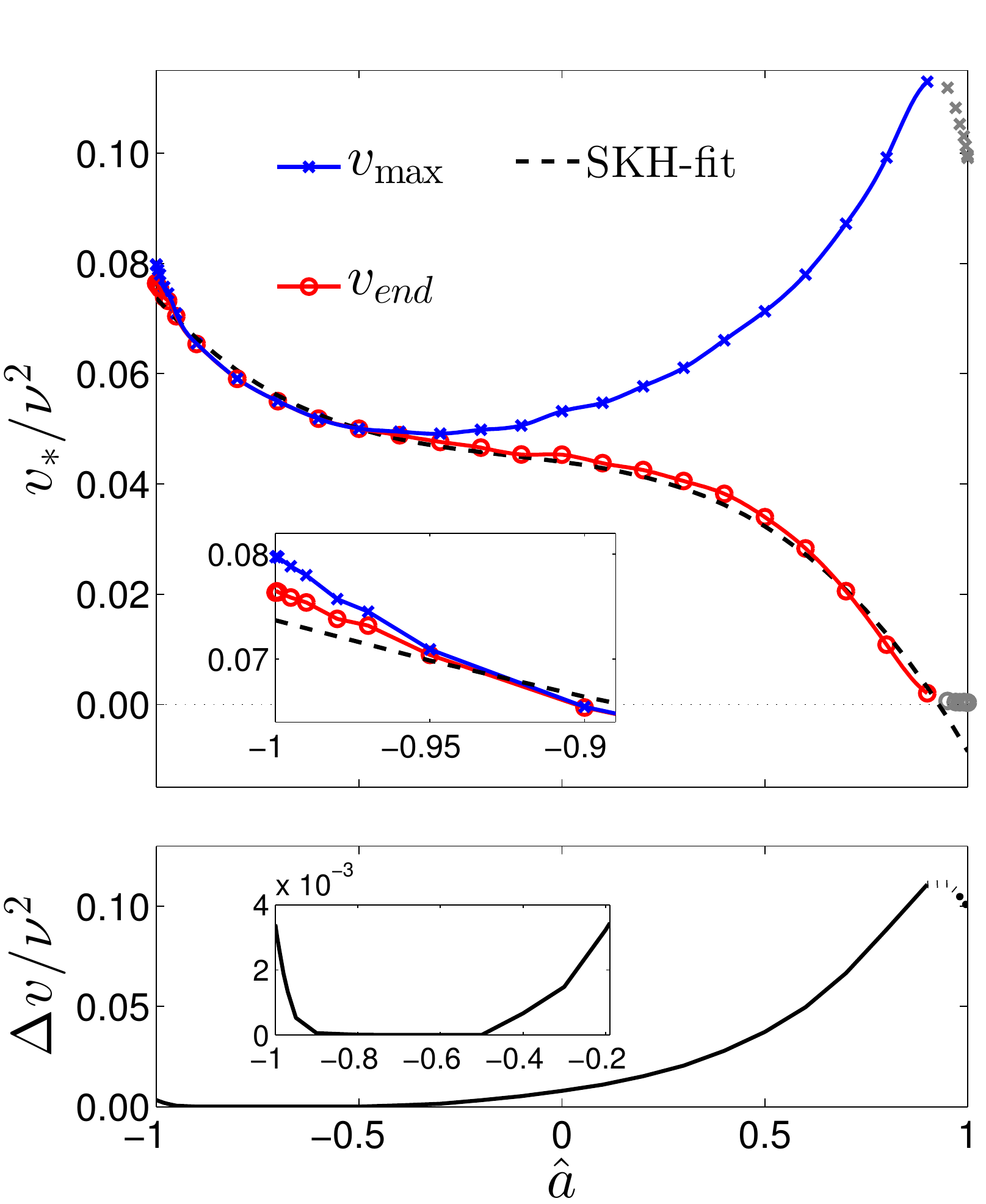} 
  \caption{Dependence of the maximum (blue, crosses) and the final (red, circles)
    recoil velocities on the spin $\ha$ for $\nu=10^{-3}$. 
    The dashed black line refers to the fit of ~\cite{Sundararajan:2010sr}.
    Although the antikick is suppressed in the interval
    $-0.9\leq\ha\leq-0.5$, it strikes back for large negative spins 
    , i.e. for $\ha\lesssim-0.9$ we find again that $v_{max} >
    v_{end}$.
    The data points for $\ha>0.9$ are plotted in gray to indicate that they
    are affected by larger systematic uncertainties due to inaccuracy of the radiation
    reaction as $\ha\rightarrow +1$ ($v_{max}$ is expected to grow monotonically. 
    See~Appendix~\ref{sec:acc_fit}).}
 \label{fig:antikick_back}
\end{figure}

Let us now discuss the recoil velocity computation and the antikick. 
We define a complex velocity vector $\mathbf{v}\equiv v_{x}+\ii v_{y}$ 
corresponding to the recoil velocity accumulated by the system up to a certain time $t$, 
\be
\label{eq:vkick}
\mathbf{v}
= -\dfrac{1}{M}\int_{-\infty}^t dt'\ \left( \F_x^{\bf P} + \ii
  \F_y^{\bf P} \right)\ .
\ee
In practice, the improper integral above is calculated from a finite initial time $t_0$. 
Thus the recoil velocity calculation requires to fix a complex integration constant
$\mathbf{v}_0$ that accounts for the velocity that the system has
acquired in evolving from $t=-\infty$ to $t = t_{0}$, i.e
\be
\label{eq:vkick_v0}
\mathbf{v} = \mathbf{v}_0
-\dfrac{1}{M}\int_{t_0}^t dt'\left( \F_x^{\bf P}+\ii \F_{y}^{\bf P}\right).
\ee
If this integration constant is not determined correctly, 
unphysical oscillations show up in the time evolution of the modulus of the velocity 
$v(t)\equiv |\mathbf{v}(t)|$, which eventually result in an inaccurate 
estimate of the final recoil.  We determine the vectorial integration 
constant $\mathbf{v}_0$  by finding the center of the hodograph of 
the velocity in the complex plane following~\cite{Pollney:2007ss,Bernuzzi:2010ty}. 
This procedure is tuned iteratively until the time evolution of $v(t)$
during inspiral grows monotonically without spurious oscillations. 
The correct determination of the integration constant is especially 
important when $\ha\to +1$, as it can strongly influence the rather small 
value of the final recoil velocity.

Figure~\ref{fig:vkick} shows for some representative configurations 
$\ha\in\{-0.9999, -0.8, -0.5, +0.2, +0.9\}$ the computed time evolution 
of the recoil velocity. Visually the ascent of the curves is  
free of oscillations due to the fine tuned setting of $\mathbf{v}_0$.
Close to merger $v(t)$ grows monotonically until
it reaches its maximum $v_{max}$. For large positive spins it then drops down
to an asymptotic value $v_{end}<v_{max}$.
The gap $\Delta v=v_{max}-v_{end}$ between the maximal and the final recoil
velocity is called the antikick. We list in Table~\ref{tab:finalV} the
values of the maximal and final recoil velocities as well as the antikick for the 
configurations considered in this work. The antikick is large for positive
spins and essentially absent for $-0.9 \leq \ha \leq -0.5$. Our data highlight
a new feature of the antikick for nearly-extremal, negative spins:
{\it the antikick ``strikes back''} for $-1<\ha<-0.9$, i.e. $\Delta v$ increases 
again, though it reaches smaller values than for positive spins. 
From the value $\Delta v/\nu^2\sim6\times 10^{{-5}}$ at $\ha=-0.9$, it 
rises to $1.3\times 10^{-3}$ at $\ha=-0.97$ and reaches $\sim3.4\times 10^{-3}$ 
in the most extremal case considered ($\ha=-0.9999$). This value is comparable 
to values obtained for $\ha\sim-0.2$.
The behavior of the recoil velocities and the antikick versus $\ha$ is illustrated in 
Figure~\ref{fig:antikick_back}. The top panel shows the maximal and final 
recoil velocities. The SKH fit is included for comparison. The bottom panel shows the
antikick. Note that in the range $-0.9\leq \ha \leq +0.9$ our data are compatible
(though different because of different accuracy, see Appendix~\ref{sec:acc_fit})
with SKH findings.

The reappearance of the antikick, although apriori surprising, can be understood 
quantitatively in relatively simple terms following DG.
One of the points of DG was to relate the antikick to
the maximum of the modulus of the GW linear momentum flux, 
$\F_{\bf P}^{max}=\max|\F_{\bf P}|$. 
At time $t$, the accumulated kick velocity is given 
by the complex integral~\eqref{eq:vkick},
i.e. $\mathbf{v}= \ii \int_{-\infty}^{t} |\F_{\bf P}(t)|
e^{i\varphi(t)}dt$, where $\varphi(t)$ is the phase of
the linear momentum flux. Expanding around the time 
$t_{max}$ corresponding to $\F_{\bf P}^{max}$ one gets 
\be
\label{eq:vt}
\mathbf{v}\simeq \ii \F_{\bf P}^{max}e^{\ii \varphi_{max}}\sqrt{\dfrac{\pi}{2\alpha}}e^{\beta^2/(2\alpha)}{\rm erfc}(z),
\ee
with $z=-\sqrt{\alpha/2}(\bar{t}-\beta/\alpha)$, where $\alpha\equiv 1/\tau_{max}^2(1-\ii\epsilon_{max})$
and $\beta=\ii Q/\tau_{max}$. Here $\tau^2_{max}\equiv -\F_{\bf P}^{max}/(\ddot{|\F|}_{\bf P})^{max}$
is the characteristic time scale associated with the ``resonance peak"
of $|\F_{\bf P}|$; $\omega\equiv \dot{\varphi}$, 
$\epsilon_{max}\equiv \dot{\omega}_{max}\tau^2_{max}$, 
and the quantity
\be
\label{eq:Q}
Q\equiv \omega_{max}\tau_{max},
\ee 
can be interpreted as the 
{\it quality factor} associated with the same peak.
According to Eq.~\eqref{eq:vt} the time evolution of the recoil
velocity is given by the complementary error function ${\rm erfc}(z)$ 
of a complex argument $z$ whose imaginary part 
is proportional to the quality factor $Q$. Hence, the quality factor $Q$
controls the monotonic behavior of ${\rm erfc}(z)$: when $Q$ is sufficiently
large a local maximum appears.

The values of $Q$ are listed in Table~\ref{tab:finalV} for all configurations
considered. One observes
immediately the tight correlations between $Q$, $v_{end}$, $\Delta v$ 
and $\tau_{\dot{p}_{r_{*}}}^{max}$, 
which supports the interpretation of the antikick results. The quantities
$\tau_{\dot{p}_{r_{*}}}^{max}$ and $Q$ behave qualitatively like $\Delta v$, 
i.e. their minima at $\ha\sim(-0.5,-0.58,-0.75$) for
($\Delta v$, $\tau_{\dot{p}_{r_{*}}}^{max}$, $Q$) are close 
and all of them increase again when $\ha\to-1$.
Physically the quality factor can be interpreted as a measure of the
adiabaticity of the process: small $Q$ indicates fast emission of 
linear momentum and reduced antikick; large $Q$ indicates
slow emission of linear momentum and enhanced
antikick. Thus, the computation of $Q$ from the maximum of the linear
momentum flux gives us a quantitative method to understand the
origin of the antikick and, in particular, to predict its behavior
for $\ha\to-1$ (see Fig.~\ref{fig:antikick_back}).
Although $Q$ is quantitative and helpful in understanding the global
picture, it might be missing some details. For example, Table~\ref{tab:finalV} 
says that $Q$ is in one to one correspondence with $\Delta v$ and 
$\tau^{\rm max}_{p_{r_{*}}}$ for all values of $\ha$ except in the range
$-0.4 < \ha < 0$, where it seems to oscillate instead of growing monotonically
as the values of $\Delta v$ suggest.
Actually, inspecting $|\F_{\bf P}|$ for, say, $\ha=-0.3$ (that shows the largest
deviation from the global growing trend) one finds that it has a rather shallow
top region, with essentially two maxima of approximately the same height 
fused together. In this particular case, the approximation that is behind the
computation of $Q$ is probably not accurate enough to faithfully represent
the structure of the peak of $|\F_{\bf P}|$ .

Finally, following DG, when $t\gg \tau_{max}$, the error function in 
Eq.~\eqref{eq:vt} can be evaluated analytically to give the final recoil magnitude
\begin{align}
\label{eq:vend}
v^{anal}_{end}\simeq \sqrt{2\pi} \F_{\bf P}^{max}\dfrac{\tau_{max}}{(1+\epsilon^2_{max})^{1/4}}e^{-Q^2/[2(1+\epsilon^2_{max})]} \quad .
\end{align}
Looking at Table~\ref{tab:finalV} the computed $v^{anal}_{end}$ is at the same order 
as $v_{end}$ over the whole spin range. Percentual differences usually vary 
around $\sim50\%$ but can reach $\sim10\%$ for values around $\ha\sim0.6$.
It would be interesting to increase
the order of the approximation of formula~\eqref{eq:vt} and recheck its
domain of accuracy depending on $\ha$. Such formula would simplify the
computation of the final recoil from numerical relativity data, especially 
because one would rely only on local knowledge of the linear momentum
flux avoiding the uncertainties related to the integration constant.

\section{Conclusions}
\label{sec:conc}

The main finding of this paper is a new phenomenon for nearly extremal
negative spins. The antikick, i.e. the
drop from the maximal to the final recoil velocity, is {\it not} a 
monotonic function of the spin and, while suppressed 
between $-0.9\leq \ha \leq-0.5$, it reappears for nearly extremal negative spins. 
Quantitatively, this surprising phenomenon is a small but significant effect, 
and its existence allows us to get a new understanding of the dynamics 
of retrograde plunges. It can be interpreted quantitatively and predicted qualitatively 
by analyzing the plunge dynamics or the GW linear momentum flux around its
maximum. The variation of the latter can be measured by the quality factor 
$Q$, which can also be viewed as a measure of the ``adiabaticity'' of the
process of emission of linear momentum through GWs. A significant antikick 
always results from a slow (quasi-adiabatic)  plunge and is associated with 
large values of $Q$. Small values of $Q$ mirror a rather nonadiabatic plunge
and, consistently, small, or absent, antikicks.

In this work we have pointed out how certain features of the linear momentum
flux directly mirror the dynamics. Qualitatively, our findings may be
robust also in unequal but comparable mass-ratio binaries, in which
the ratio between the spin of the two objects is nearly extremal. 
The flux analysis presented here may guide the extraction of
useful information for kick computations in numerical relativity, 
like those recently performed in~\cite{Healy:2014yta}.

\begin{acknowledgments}
This work was supported in part by  DFG grant SFB/Transregio~7
``Gravitational Wave Astronomy''. 
E.H., S.B., and A.Z. thank IHES for hospitality during the development of part of this work.
A.N. acknowledges Thibault Damour for useful discussions.
\end{acknowledgments}

\appendix
\section{Accuracy}
\label{sec:acc_fit}

\begin{table}[t]
  \centering
    \caption{Dependence on $\ell_{max}$ of $v_{end}$ and $v_{max}$.
    For $\ha=-0.9999$, $\ell_{max}>4$ contributions give less than
    $1\%$. For $\ha=+0.9$, the effect is larger and $v_{end}$
    slightly increases for higher $\ell_{max}$.}
    \label{tab:importance_lmax} 
    \begin{tabular}[t]{c||cccc}
      \hline
      \multicolumn{5}{c}{$\ha=-0.9999$} \\
      \hline
      $\ell_{max}$ & $v_{max}/\nu^2$ & diff $[\%]$ & $v_{end}/\nu^2$ & diff $[\%]$  \\
      \hline
      2 & 0.070252 &  -    & 0.068323 &  -    \\
      3 & 0.077692 & 10.59 & 0.074520 &  9.07 \\
      4 & 0.079033 & 1.73  & 0.075589 &  1.43 \\
      5 & 0.079187 & 0.19  & 0.075766 &  0.23 \\
      6 & 0.079442 & 0.32  & 0.076045 &  0.37 \\
      7 & 0.079613 & 0.21  & 0.076228 &  0.24 \\
      8 & 0.079722 & 0.14  & 0.076345 &  0.15 \\
      \hline
    \end{tabular}     
    \begin{tabular}[t]{c||cccc}
      \hline
      \multicolumn{5}{c}{$\ha=+0.9$} \\
      \hline
      $\ell_{max}$ & $v_{max}/\nu^2$ & diff $[\%]$ & $v_{end}/\nu^2$ & diff $[\%]$  \\
      \hline   
      2 & 0.003687 &  -      &  0.000932 &  -     \\
      3 & 0.045957 & 1146.49 & 0.001190  &  27.80 \\
      4 & 0.074009 & 61.04   & 0.001350  &  13.43 \\
      5 & 0.091701 & 23.90   & 0.001535  &  13.64 \\
      6 & 0.102239 & 11.49   & 0.001741  &  13.44 \\
      7 & 0.108800 & 6.42    & 0.001917  &  10.10 \\
      8 & 0.112927 & 3.79    & 0.002056  &  7.27  \\
      \hline
    \end{tabular}     
\end{table}

\begin{table}[t]
  \centering
    \caption{Effect of the mass ratio $\nu$. The table
    compares for a few values of $\ha$ the recoil 
    velocities as obtained from trajectories with $\nu=10^{-3}$
    and $\nu=10^{-4}$. The percentual difference if about $1\%$ for
    $\ha<0.9$ and reaches
    $\sim7\%$ for $\ha=0.9$. We use the notation $v^{(\log_{10}\nu)}$.}
    \label{tab:importance_mue} 
    \begin{tabular}[t]{c||ccc|ccc}
      \hline
      $\ha$ & $v_{max}^{(-3)}/\nu^2$ & $v_{max}^{(-4)}/\nu^2$ & diff
      $[\%]$ &  $v_{end}^{(-3)}/\nu^2$ & $v_{end}^{(-4)}/\nu^2$ & diff
      $[\%]$   \\ 
      \hline
      -0.9000 & 0.06545 & 0.06598 & 0.81 & 0.06539 & 0.06592 & 0.81 \\
      -0.7000 & 0.05501 & 0.05504 & 0.06 & 0.05501 & 0.05504 & 0.06 \\
      -0.5000 & 0.05003 & 0.04964 & 0.76 & 0.05003 & 0.04964 & 0.76 \\
      0.0000 & 0.05319 & 0.05313 & 0.11 & 0.04530 & 0.04508 & 0.48 \\
      0.5000 & 0.07131 & 0.07119 & 0.17 & 0.03398 & 0.03383 & 0.44 \\
      0.7000 & 0.08719 & 0.08877 & 1.81 & 0.02056 & 0.02073 & 0.83 \\
      0.9000 & 0.11293 & 0.12093 & 7.09 & 0.00206 & 0.00199 & 3.24 \\
      \hline
    \end{tabular}     
\end{table}

We give here some estimates about the accuracy of our
computation and discuss the limitations of our approach for
configurations with $\ha\to+1$.

Table~\ref{tab:importance_lmax} shows the effect of
$\ell_{max}$ on the velocity computation. The results for
$\ha=-0.9999$ vary $\lesssim 1\%$ by including multipoles with $\ell_{max}>4$. The
inclusion of high multipoles is more relevant for large positive
spins. For $\ha=+0.9$ we observe a $\sim7\%$ variation by increasing
$\ell_{max}=7$ to $\ell_{max}=8$. Including only up to $\ell_{max}=6$
multipoles underestimates $v_{end}$ by at least $10\%$. This is consistent
with the corresponding variations we see in the fluxes, Fig.~\ref{fig:Pflux}.
Note that $v_{end}$ increases by including more multipoles.

Another source of uncertainty is the finite value of the mass ratio $\nu$ employed 
in the simulations~\cite{Bernuzzi:2010ty}. Table~\ref{tab:importance_mue}
shows a comparison between results obtained with $\nu=10^{-3}$ and
$\nu=10^{-4}$. The uncertainties for $\ha<0.5$ are at the $1\%$
level. For larger spins they grow and reach about $7\%$ for $\ha=0.9$. 
We expect even larger uncertainties for $\ha\geq0.95$ since these simulations 
are strongly biased by the inaccurate 
radiation reaction (see below). 

Our kick calculation in Table~\ref{tab:finalV} and Fig.~\ref{fig:antikick_back}
can be compared with the fit proposed in SKH. The latter was calculated
(i) including multipoles up to $m_{max}=6$; (ii) using a different technique
to determine the integration constant; and (iii) using $\nu=10^{-4}$ 
simulations of about $25$ orbits. The fit of SKH refers to the interval $|\ha|<0.9$
and is therein consistent with our data, in some cases within $1\%$. However, it does 
not capture the fine structures for nearly-extremal values of the spin. 
Observe, for example, that it underestimates $v_{end}$ for $\ha\to-1$
(Fig.~\ref{fig:antikick_back}).

Let us finally discuss the data for nearly-extremal positive spins $+0.9 < \ha \leq + 0.9999$.
These data are displayed in Fig.~\ref{fig:antikick_back} in gray color since they are 
uncertain. The numbers behind the plot are listed 
in Table~\ref{tab:final_extremal}. Inspecting the table and Fig.~\ref{fig:antikick_back} 
one sees that: (i)  $v_{end}$ first decreases and then remains approximately
constant (and very small) for $\ha \geq 0.995$; (ii) $v_{max}$ {\it decreases} monotonically;
(iii) $Q$ oscillates around 9.2 for $0.99\leq \ha \leq 0.9999$; (iv) $\tau^{max}_{\dot{p}_{r_{*}}}$
increases monotonically.
At first sight these numbers look contradictory. The increase 
of $\tau^{max}_{\dot{p}_{r_{*}}}$ with $\ha$ is indicating that the dynamics (and thus the 
emission of linear momentum) is increasingly adiabatic as $\ha\to +1$. Consistently,
$v_{end}$ decreases, but the increased adiabaticity of the dynamics is not mirrored
in $Q$ nor in $v_{max}$, which decreases instead.

\begin{table}[t]
 \caption{\label{tab:final_extremal} Same as Table~\ref{tab:finalV} for$+0.95\leq \ha\leq +0.9999$.}
  \centering
  \begin{ruledtabular}
  \begin{tabular}{l||ccc|ccc}
    $\ha$ & $v_{max}/\nu^{2}$ & $v_{end}/\nu^{2}$  &
    $\Delta v /\nu^{2}$ &  $Q$ & $\tau^{max}_{\dot{p}_{r_{*}}}$ & $v_{end}^{anal}$\\
    \hline 
    \hline   
	0.9500 & 0.11186 & 0.00065 & 1.112e-01 & 7.1404 & 8.6964 & 0.00015 \\
        0.9700 & 0.10821 & 0.00046 & 1.077e-01 & 7.9190 & 8.8428 & 0.00008 \\
        0.9800 & 0.10524 & 0.00043 & 1.048e-01 & 8.7525 & 9.0199 & 0.00021 \\
        0.9900 & 0.10307 & 0.00044 & 1.026e-01 & 9.2251 & 9.4295 & 0.00045 \\
        0.9950 & 0.10127 & 0.00038 & 1.009e-01 & 9.3933 & 9.8429 & 0.00039 \\
        0.9990 & 0.09968 & 0.00036 & 9.933e-02 & 9.2492 & 10.4124 & 0.00019 \\
        0.9999 & 0.09914 & 0.00035 & 9.878e-02 & 9.1388 & 10.5938 & 0.00031 \\   
  \end{tabular} 
  \end{ruledtabular}
\end{table}

\begin{figure}[t]
  \centering
   \includegraphics[width=0.45\textwidth]{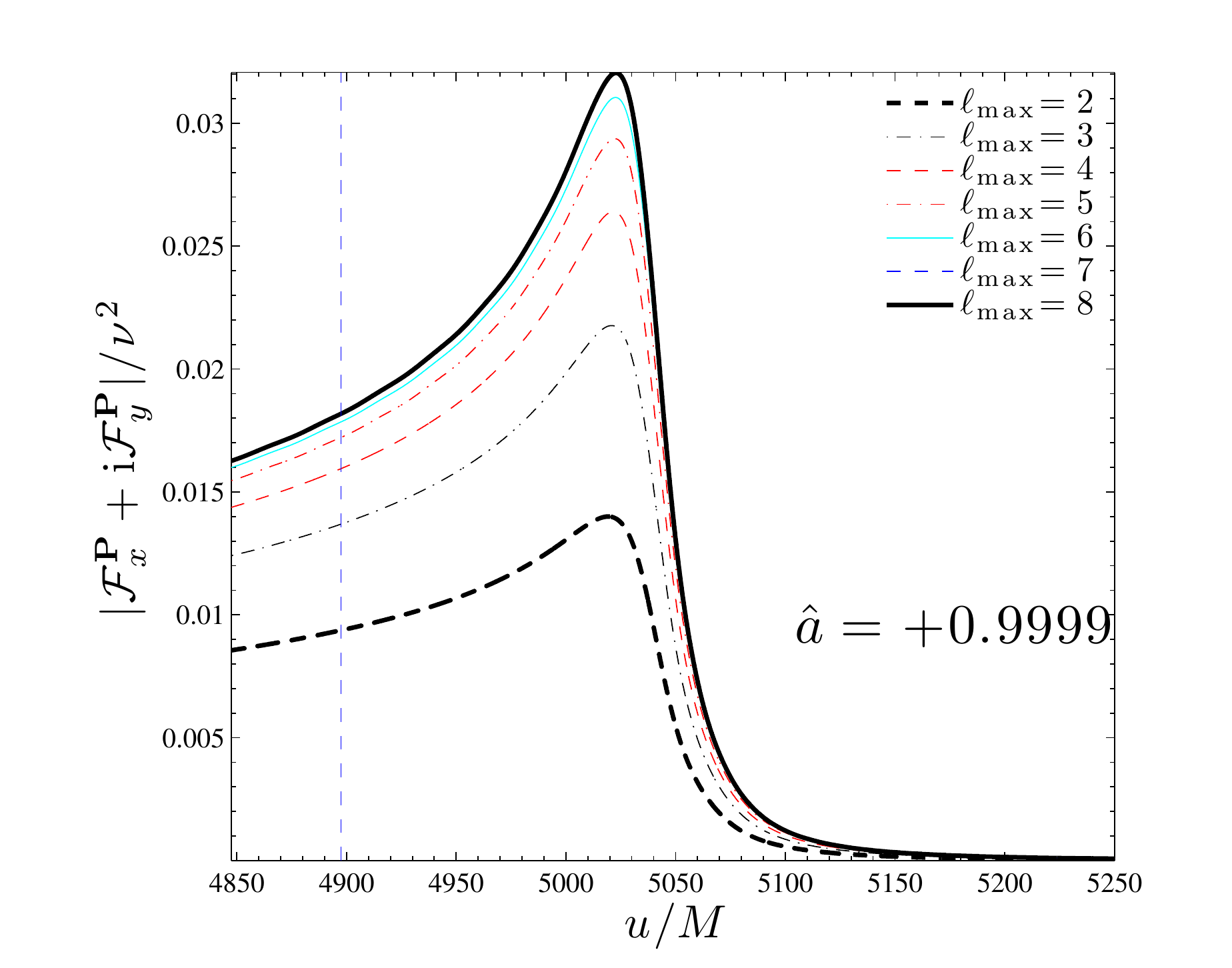} 
  \caption{Flux of linear momentum for $\ha=+0.9999$. The vertical line indicates the peak of 
  $|\Psi_{22}|$.}
 \label{fig:flux_09999}
\end{figure}

A careful inspection of the dynamics brought us to conclude that these results are 
{\it qualitatively inaccurate} for $v_{max}$ (and thus $Q$) and {\it quantitatively inaccurate}
for $v_{end}$. The main reason is the systematic inaccuracy of the
radiation reaction for large positive spins $\ha\gtrsim 0.9$, as shown in~\cite{Harms:2014dqa}.
Practically speaking, the low accuracy of the radiation reaction (and in particular the absence of
horizon fluxes that could contrast the loss of angular momentum to infinity via 
superradiance~\cite{Bernuzzi:2012ku,Taracchini:2013wfa}) makes the system lose
too much angular momentum. For $\ha > +0.97$ this effect is
so strong that the angular momentum $p_{\phi}$ becomes {\it negative}  ($p_{\phi}\sim -0.1$) 
around merger. For example, for $a+0.9999$ (see Fig.~\ref{fig:flux_09999})
this change of sign occurs at $t/M=5038$, that is quite close to the peak of the flux of 
linear momentum in a domain where the waveforms are still influenced by the dynamics 
(the LSO is crossed at $t/M=5056.6$ and the light-ring at $t/M=5220$).
This unphysical effect  ($p_{\phi}$ is defined to be positive) 
mirrors an excessive acceleration of the dynamics during the plunge 
and heuristically explains the drop of $v_{max}$ for $\ha>0.9$. 
By contrast, we found  that the calculation of $\tau^{max}_{\dot{p}_{r_{*}}}$ 
relies on a part of the dynamics before the change of sign of $p_{\phi}$ 
($-\dot{p}_{r_{*}}$ peaks at $t/M=5028$) and therefore is more robust, 
as confirmed by the monotonic behavior of $\tau^{max}_{\dot{p}_{r_{*}}}$
over $\ha$.
A way of treating larger spin values is to adopt the self-consistent radiation
reaction method introduced in~\cite{Harms:2014dqa}. Doing this is
computationally very demanding and will be
discussed in a follow up study. At present, we could check our understanding
only against self-consistent $\ha=+0.9$ data~\cite{Harms:2014dqa}.
Consistently with our expectation that the correct radiation reaction should
yield a more adiabatic plunge, we found a slightly smaller 
$v_{end}^{\rm sc}/\nu^{2}=0.00189$  (instead of 0.00206) and a slightly 
larger $v_{max}^{\rm sc}/\nu^{2}=0.11908$ (instead of 0.11293).
This preliminary result suggests that $v_{max}$ will increase further and
$v_{end}$ will become smaller as $\ha\to +1$. New,
challenging investigations will be needed to assess whether 
$v_{end}= 0$ as $\ha=+1$.


\end{document}